 \documentclass[10pt,journal,epsfig]{IEEEtran}

\usepackage[dvips]{graphicx}
\usepackage{graphicx}
\usepackage{amssymb}
\usepackage{cite}
\usepackage{subfigure}
\usepackage{amsmath}
\usepackage{algorithm}
\usepackage{algorithmic}
\usepackage{multirow}
\usepackage[table]{xcolor}

\usepackage{makecell}
\usepackage{diagbox}
\usepackage{array}
\usepackage{threeparttable}

\begin{document}

\title{Beam Training and Alignment for RIS-Assisted Millimeter Wave Systems:
State of the Art and Beyond}

\author{Peilan Wang, Jun Fang, Weizheng Zhang, Zhi Chen,
Hongbin Li,~\IEEEmembership{Fellow,~IEEE}, and Wei
Zhang,~\IEEEmembership{Fellow,~IEEE}
\thanks{Peilan Wang, Jun Fang and Zhi Chen are with the University of
Electronic Science and Technology of China. Email:
201811220519@std.uestc.edu.cn; JunFang@uestc.edu.cn;
chenzhi@uestc.edu.cn}
\thanks{Weizheng Zhang and Wei Zhang are with the University of New South
Wales. Email: zwz@ieee.org; w.zhang@unsw.edu.au}
\thanks{Hongbin Li is with the Stevens Institute of Technology. E-mail:
Hongbin.Li@stevens.edu}
\thanks{This is an open call article which is recommended for
acceptance by Dr. Xiaojiang Du.}
\thanks{This work was supported in part by the National Science
Foundation of China under Grant 61829103, in part by the National
Key R\&D Program of China under Grant 2018YFB1801500, the Key Area
Research and Development Program of Guangdong Province under grant
No. 2020B0101110003, and in part by Shenzhen Science \& Innovation
Fund under Grant JCYJ20180507182451820..\copyright  2022 IEEE.  Personal use of this material is permitted.  Permission from IEEE must be obtained for all other uses, in any current or future media, including reprinting/republishing this material for advertising or promotional purposes, creating new collective works, for resale or redistribution to servers or lists, or reuse of any copyrighted component of this work in other works.} }

\maketitle

\begin{abstract}
Reconfigurable intelligent surface (RIS) has recently emerged as a
promising paradigm for future cellular networks. Specifically, due
to its capability in reshaping the propagation environment, RIS
was introduced to address the blockage issue in millimeter Wave
(mmWave) or even Terahertz (THz) communications. The deployment of
RIS, however, complicates the system architecture and poses a
significant challenge for beam training (BT)/ beam alignment (BA),
a process that is required to establish a reliable link between
the transmitter and the receiver. In this article, we first review
several state-of-the-art beam training solutions for RIS-assisted
mmWave systems and discuss their respective advantages and
limitations. We also present a new multi-directional BT method,
which can achieve a decent BA performance with only a small amount
of training overhead. Finally, we outline several important open
issues in BT for RIS-assisted mmWave systems.
\end{abstract}


\begin{figure*}[!t]
\centering
\includegraphics[width=7in] {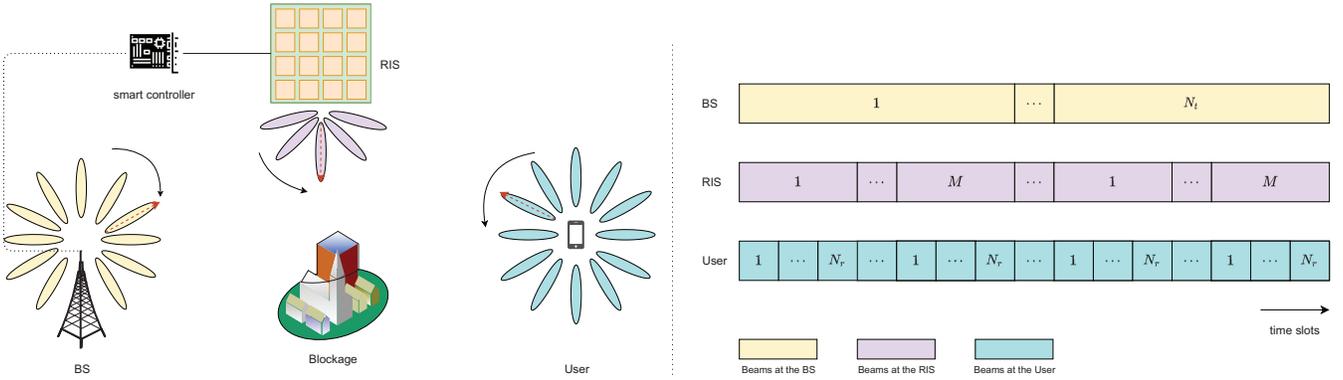}
\caption{Exhaustive search for RIS-assisted mmWave downlink
systems, where a BS with $N_t$ antennas serves the user with $N_r$
antennas aided by an RIS, which is a planar array consisting of $M
= M_y \times M_z$ passive reflecting elements. The exhaustive
search scheme needs $N_tMN_r$ time slots to perform BT.
} \label{Fig_exhaus}
\end{figure*}

\section{Introduction}
Millimeter-wave (mmWave) communications have been considered as an
enabling technology for beyond 5G (B5G) or 6G systems to
accommodate the exponentially increasing demand of wireless data
services. Nevertheless, mmWave signals suffers severe path loss,
e.g., 100dB/100m at 28GHz \cite{AkdenizLiu14,RappaportXing19}. To
compensate for such path loss, large antenna arrays and
beamforming techniques are utilized at the transceiver to provide
adequate link budgets. Directional beamforming, however, makes
mmWave communications vulnerable to blockage. For instance, it was
reported in \cite{AbariBharadia17} that the mmWave link can be
easily blocked by a small obstacle such as a person's arm. Hence,
blockage is considered a major issue that hampers the deployment
of mmWave communications in cellular networks.

To alleviate the blockage problem, reconfigurable intelligent
surface (RIS), a.k.a., metasurface and large intelligent surface,
has recently emerged as a promising paradigm to establish a
virtual line-of-sight (LOS) path when the direct link is blocked
by obstructions \cite{WangFang19,TanSun18}. Specifically, RIS is a
planar surface consisting of a massive number of low-cost and
passive reflecting units. Each unit or element can induce an
adjustable amplitude and phase shift to reflect the incident
electromagnetic waves with the aid of a smart controller
\cite{LiaskosNie18}. RIS-assisted systems can thus help reshape
the wireless propagation environment via software-controlled
reflection.

Recent progress has shown the great potential of RIS as an
auxiliary device in boosting spectral efficiency, mitigating
interference, enhancing physical security and so on
\cite{WuZhang21}. In particular, since RIS only passively reflects
the impinging signals, it can be operated in an energy-efficient
way without the need of radio frequency (RF) chains, and can thus
reduce the energy consumption by orders of magnitude compared with
traditional active antenna arrays. Also, due to its passive
characteristics, RIS is free of self-interference and antenna
noise amplification. Recent theoretical analyses reveal that
RIS-assisted systems can achieve a quadratic scaling law in the
receive signal power which scales quadratically with the amount of
passive units \cite{WuZhang21,WangFang19}. All these amiable
features make RIS an appealing solution for overcoming blockage
and improving coverage in mmWave communications.

Channel state information (CSI) acquisition is a pre-requisite to
realize the full potential of RIS-assisted mmWave systems.
Nevertheless, obtaining the \emph{full CSI} of the BS-RIS (B-R)
link and the RIS-user (R-U) link is very challenging since RIS
cannot receive or transmit signals and lacks signal processing
capabilities. Moreover, the CSI acquisition requires a large
amount of training overhead because of the giant size of antenna
arrays at the transceiver as well as the large number of passive
elements at the RIS. It should be noted that some recent efforts,
e.g. \cite{WangFang20,WeiShen21,WanGao21}, attempted to exploit
the inherent sparse nature of the cascade BS-RIS-user mmWave
channel and formulate the channel estimation problem into a
compressed sensing (CS) framework. CS-based methods, however, are
expensive to implement due to its excessive computational
complexity.

On the other hand, many channel measurements \cite{AkdenizLiu14}
indicated that mmWave channels exhibit sparse scattering
characteristics. Particularly, it was reported that the power of
the LOS path is much higher (about 13dB higher in mmWave bands and
20dB higher in THz bands) than the sum of the power of
non-line-of-sight (NLOS) paths. Therefore, instead of obtaining
the full CSI, an alternative option is to identify only the
dominant path, and align the transmitter's and receiver's beams to
provide a sufficient beamforming gain for mmWave communications.
The procedure devised for identifying one or several dominant
paths for initial access is referred to as beam training (BT) or
beam alignment (BA). For RIS-assisted mmWave systems, BT aims to
simultaneously identify the best BA for both the B-R link and the
R-U link. As RIS is a passive device that cannot transmit/receive
signals, BT for RIS-assisted mmWave systems is more challenging
than that for conventional mmWave systems.

In this article, we discuss the extension of conventional BT
methods to RIS-assisted mmWave systems, and elaborate the pros and
cons of each approach. We also propose a novel multi-directional
BT scheme for RIS-assisted systems to address drawbacks of
existing BT methods. Numerical results are provided to show the
efficiency of the proposed solution. Finally, future research
challenges are discussed, followed by concluding remarks.

\section{Beam Training for RIS-Assisted MmWave Systems}
We focus on the scenario where the BS-user link is blocked owing
to unfavorable propagation environments. To perfectly align the
beams for the three-node communication system, BT involves
estimating the angle of departure (AoD) and the angle of arrival
(AoA) associated with the dominant path of the B-R link, and the
AoD and the AoA associated with the dominant path of the R-U link.
Due to the cascade nature of the channel, by carefully devising
its phase shift vector, RIS can form a reflecting beam pointing in
a direction that is a superposition of the incident angle and a
``relative reflection'' angle. Such a carefully devised phase
shift vector takes a form of conventional steering vectors and can
be characterized by the relative reflection angle (RRA).
Therefore, instead of searching for the AoA and AoD at the RIS, we
aim to find a best RRA at the RIS to achieve BA for both the B-R
and the R-U links. In doing this way, the search space can be
significantly reduced.

In practice, with the knowledge of the location of the BS, RISs
can be installed within the sight of the BS. Hence, some existing
works, e.g. \cite{YouZheng20}, assume that the BS has aligned its
beam to the RIS, and focus on the BT between the RIS and the user,
which can be accomplished by using conventional BT techniques.
Nevertheless, the above assumption is only valid for stationary
BSs and RISs. With the popularity of unmanned aerial vehicles
(UAVs) and their promising potential in wireless communications,
mobile BSs based on UAVs and the likes are being considered for
possible deployment in the near future. In such scenarios, it is
necessary to simultaneously identify the best BA for both the B-R
link and the R-U link. Moreover, in practice, the LOS path between
the BS and the RIS may be blocked by obstacles, in which case we
also need to launch joint BS-RIS-user BT to find an alternative
path from the BS to the user. In the following, we first discuss
the extension of conventional BT techniques to RIS-assisted mmWave
systems.

\subsection{Exhaustive Search}
A natural approach to perform BT is to exhaustively search all
possible beam tuples/triplets. Specifically, the BS, the RIS and
the user adopt pre-designed codebooks, $\mathcal{F}_{B}$,
$\mathcal{F}_{R}$ and $\mathcal{F}_{U}$ respectively. Each
codebook, consisting of a set of narrow beams, is designed by
uniformly quantizing the associated beam angle, i.e. the AoD (AoA)
for the BS (user) and the RRA for the RIS. The finely quantized
angles are assumed to uniformly cover the whole range of the
AoD/AoA/RRA angles. The best beam tuple for BA is identified by
exhaustively searching all possible $| \mathcal{F}_B|
|\mathcal{F}_R| |\mathcal{F}_U|$ beam tuples based on the received
signal power. Such an exhaustive search requires the RIS scans its
entire RRA angular space for each choice of beam direction on the
BS side, and meanwhile requires the receiver (i.e. user) to scan
its entire AoA space for every combination of AoD and RRA, as
illustrated in Fig. \ref{Fig_exhaus}.

Due to the use of pencil beams, an important advantage of the
exhaustive search scheme is that the BS-RIS-user link achieves a
large beamforming gain if the beam tuple aligns with the dominant
path, which can help identify the best BA even in low
signal-to-noise ratio (SNR) environments. The exhaustive search
method, albeit simple and straightforward, entails a prohibitively
large training overhead. Suppose the BS (user) is equipped with
$N_t$ ($N_r$) antennas, and the RIS is a planar surface with $M =
M_y \times M_z$ reflecting elements. To achieve a highest spatial
resolution, the BS (user) needs to use $N_t$ ($N_r$) narrow beams
to scan the entire AoA (AoD) space, and the RIS needs to use $M$
reflecting beams to search the RRA domain. The total number of
beam tuples to be examined is therefore up to $N_t N_r M$. Let
$N_t = 32$, $N_r = 32$ and $M = 256$. Then the number of possible
beam combinations is $262144$, which would result in an excessive
delay for initial access.

\begin{figure}[!t]
\centering
\includegraphics[width=3.5in] {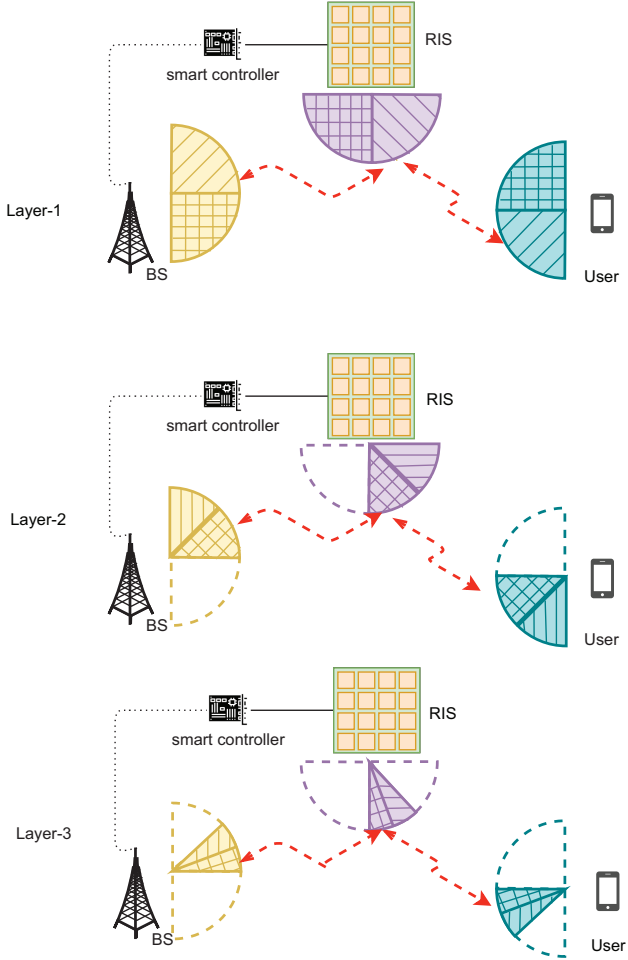}
\caption{A $3$-layer hierarchical search with $3 \times 2 \times 2
\times 2 = 24$ beam tuples for RIS-assisted mmWave downlink
systems.} \label{Fig_hier}
\end{figure}

\subsection{Hierarchical Search}
To expedite the BA process, hierarchical multi-resolution beam
search approaches were proposed for conventional mmWave systems
\cite{XiaoHe16}. The hierarchical search scheme
can be readily extended to the RIS-assisted systems, where the BS,
the RIS and the user employ their respective multi-layer
beamforming codebooks for joint spatial scanning. For hierarchical
codebooks, a lower-layer codebook consists of wider beams in
comparison with higher-layer codebooks. The spatial resolution
rises as the number of layers increases. The hierarchical search
scheme consists of multiple stages. At each stage, the
corresponding layer's codebooks/subcodebooks are used for spatial
scanning. The scanning procedure is similar to that of the
exhaustive search scheme, except that here we only need to search
the range that is identified in the earlier stage. The user then
examines the received signal power to find the best beam tuple.
This information is fed back to the BS and RIS to adaptively
choose higher-resolution subcodebooks for the next stage's
scanning. The above process continues until the required spatial
resolution is achieved, which is depicted in Fig.\ref{Fig_hier}.

The hierarchical search can effectively reduce the amount of
training overhead. Considering a typical binary-tree search. At
each stage, the search range at each node is halved by two beams
and the required number of layers for hierarchical search is
determined by $\log_2 \max\{N_r,M,N_t\}$. Let $N_t=32$, $N_r =
32$, and $M =256$. We will need $\log_2 256=8$ layers to complete
the hierarchical search. There are $2^3=8$ beam tuples to be
examined for each of the first five stages. For the last three
stages, the beams at the BS and the user are fixed, and we only
need to search the RRA at the RIS. Thus the total number of beam
tuples to be examined throughout the whole process is given by
$(5\times 8) + (3 \times 2) = 46$, which is far less than that
required by the exhaustive search scheme.

Despite the substantial training overhead reduction, a major
drawback of hierarchical beam search is that the use of wide beams
at early stages results in low beamforming gains. As a result,
spatial scanning at lower levels may fail to identify the correct
beam tuple in low SNRs and eventually miss detecting the dominant
path. The situation gets worse for RIS-assisted mmWave systems due
to the severe product-path-loss of the cascade channel. Also, the
hierarchical beam search scheme involves frequent feedback from
the user to the BS/RIS, which exerts an extra burden on the
training process. Lastly, since hierarchical beam search needs the
BS/RIS interact with each user individually, the extension to
multi-user scenarios requires careful global coordination,
presenting another challenge for such systems.

\subsection{Other State-of-The-Art Methods}
In addition to the exhaustive search and hierarchical search
schemes, there have been some attempts that utilize recent
progress in machine learning (deep neural networks, reinforcement
learning and so on) to tackle the mmWave BT problem, e.g.
\cite{HengAndrews21}. This type of machine learning-based methods
can also be extended to RIS-assisted mmWave systems. On the other
hand, although machine learning-based methods have the potential
to achieve superior performance, they often require massive amount
of data and rich computational resources for offline training. It
is certainly worth further investigation to find out their pros
and cons as compared with the conventional BT methods.

It was also brought to our attention that MUSIC algorithms have
been recently applied to mmWave MIMO systems to obtain the angular
parameter information \cite{WeiZhou17}. The basic idea behind
MUSIC algorithms is to exploit the geometric structure of the
array manifold and the phase difference of received signals on
different antennas. Therefore MUSIC-type methods need accurate
phase information of the received signals. In contrast, most BT
schemes (including our proposed method) only utilize the magnitude
information of the received measurements for angle estimation.
Note that in mmWave bands, due to the carrier frequency offset
(CFO) effect and random phase noise, we may have to rely only on
the magnitude of the measurements for BA at the initial access
stage.

\section{Multi-Directional Beam Training Scheme}
To overcome difficulties faced by existing BT approaches, we, in
this section, propose a multi-directional BT scheme for
RIS-assisted mmWave systems by leveraging the sparse
characteristics of mmWave channels. The idea is to let each node
in the system simultaneously form multiple directional beams to
probe its associated angular (i.e., AoA, AoD, RRA) space.
Specifically, the angular space can be divided into a number of
disjoint spatial intervals via beams formed by a discrete Fourier
transform (DFT) matrix. To develop a multi-directional BT scheme
for RIS-assisted mmWave systems, we first discuss how to
simultaneously generate multiple DFT beams at each node of the
system.

\subsection{Generating Multi-Directional Beams}
In mmWave systems, a hybrid analog and digital
beamforming/combining structure is usually employed at the BS/user
to reduce the cost/energy consumption. We assume a fully-connected
(FC) hybrid structure is used at the BS/user, where each RF chain
is connected to all antennas. For a fully-connected hybrid
structure, the BS (user) can simultaneously form $R_{\text{BS}}$
($R_{\text{UE}}$) directional beams by simply letting the hybrid
precoder (combiner) be a linear combination of $R_{\text{BS}}$
($R_{\text{UE}}$) columns of an $N_t\times N_t$ ($N_r\times N_r$)
DFT matrix. Here $R_{\text{BS}}$ ($R_{\text{UE}}$) denotes the
number of RF chains at the BS (user).

At the RIS, we can generate a number of reflecting beams
simultaneously by setting the reflection vector to be a linear
combination of $Q$ distinct columns of an $M\times M$ DFT matrix.
Here $Q$ is a parameter of user's choice. Such an approach,
however, requires an independent control of each RIS element's
reflection amplitude. To address this difficulty, an alternative
solution is to generate multiple DFT beams using a phase shift
vector, which can be cast into a constrained optimization problem
\cite{WangFang21a}.

\subsection{Best Beam Tuple Estimation}
Identifying the best beam tuple is equivalent to determining BS's
AoD, user's AoA, and RIS's RRA associated with the dominant path.
To this objective, one needs to search for the three-dimensional
AoD-AoA-RRA space. We can imagine this three-dimensional angular
space as a three-dimensional cube which is uniformly divided into
$N_t N_r M$ small blocks, where each block corresponds to a
possible beam tuple, and can also be viewed as a potential
BS-RIS-user path. By simultaneously generating multiple DFT beams
at each node of the system, the user collects signals coming from
multiple (more precisely, $R_{\text{BS}} R_{\text{UE}} Q$) blocks
(i.e. paths). If the dominant path matches one of these
$R_{\text{BS}} R_{\text{UE}} Q$ blocks, the user will receive a
prominent measurement. Note that the best beam tuple estimation is
performed at the receiver (user) side. After the best beam tuple
is identified, this information will be fed back to the BS via a
dedicated channel for subsequent joint beamforming and downlink
data transmission.

Specifically, we divide the three-dimensional cube into $S=N_t N_r
M/(R_{\text{BS}} R_{\text{UE}} Q)$ disjoint subsets, also referred
to as bins, each of which consists of $R_{\text{BS}} R_{\text{UE}}
Q$ blocks. Thus we can complete scanning the entire
three-dimensional angular space using $S$ successive time slots.
Such a scanning is termed as \emph{a round of batch-mode
scanning}. After a round of batch-mode scanning, we are able to
identify which bin contains the dominant path. Nevertheless, we
cannot determine which block in this bin is associated with the
dominant path.

To identify the exact block associated with the dominant path, we
perform a few rounds of batch-mode scanning, and for each round,
we randomize the blocks that fall into $S$ disjoint bins. For each
round of scanning, say, the $l$th round of scanning, we denote the
bin which contains the dominant path as $\mathcal{B}_l$. Then we
can retrieve the block associated with the dominant path by
finding the common element (i.e. block) among the bins
$\{\mathcal{B}_l\}_{l=1}^L$. The rationale behind this
intersection scheme is as follows: since the blocks assigned to
each bin at each round of scanning are randomized, it is very
unlikely that, other than the block associated with the dominant
path, there exists another block which lies in the intersection of
these bins, particularly when $L$ is large. Therefore, it is
expected that we can identify the block associated with the
dominant path with a high probability.

\begin{figure*}[!t]
\centering
\includegraphics[width=7in] {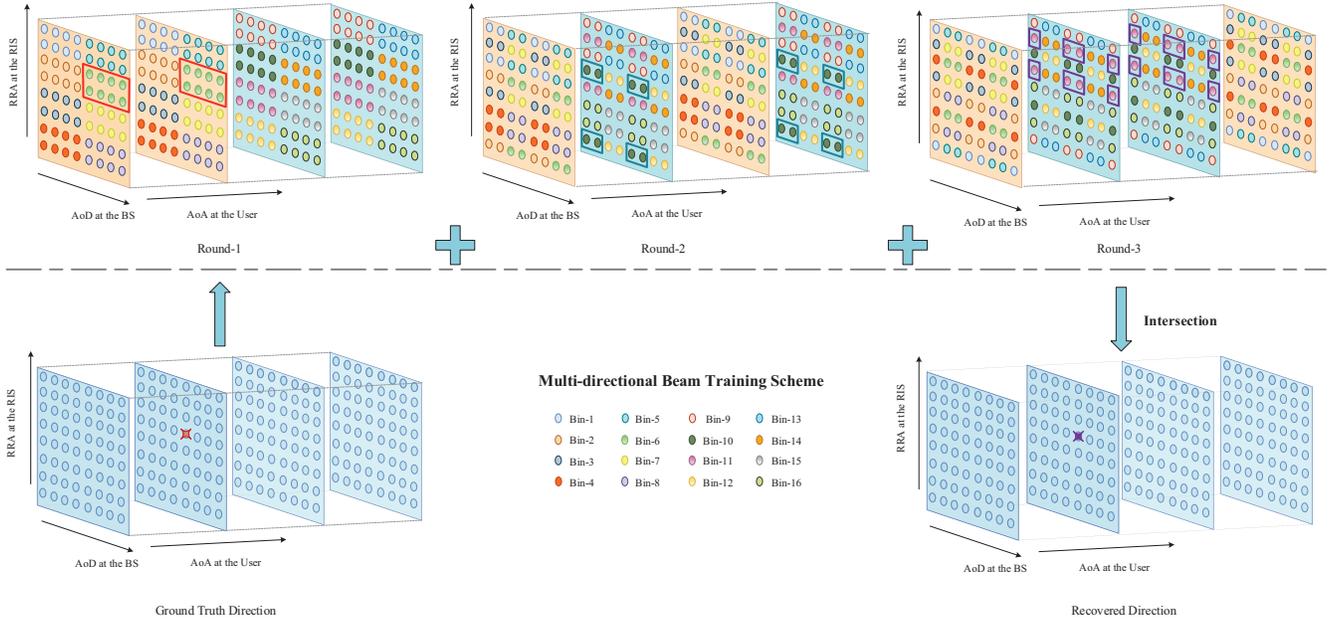}\hfill
\caption{An illustrative example to show how the proposed scheme
identifies the dominant path for RIS-assisted MIMO systems. Each
circle ``$\circ$'' represents a block. A bin consists of multiple
blocks which are sensed at a certain time slot via
multi-directional beams. For different rounds, we randomize the
blocks that fall into different bins. In each round, the subset
$\mathcal{B}_l$ is highlighted with a rectangular box. By
performing the intersection procedure, the block with the dominant
path can be estimated as the common element of those rectangular
boxes.} \label{Fig_multi-directional}
\end{figure*}

In Fig. \ref{Fig_multi-directional}, we provide an illustrative
example to show how the proposed method identifies the dominant
path in an efficient way. Consider a toy example where we have
$N_t = 8$, $N_r = 4$, $M=8$, $R_{\rm BS} = 4$, $R_{\rm UE}=2$ and
$Q=2$. The dominant BS-RIS-user path is associated with the block
located at position $(5,4,2)$. The proposed method starts by
dividing the three-dimensional space into $S = 16$ disjoint bins.
Each bin thus collects energy from paths associated with its
blocks. By examining the received signal power at the user, in
each round we can identify the bin which contains the dominant
path. For instance, in round $1$, bin $6$ is identified and
highlighted with a red rectangular box, bin $10$ is marked in the
second round and so on. Finally, we can identify the dominant path
by finding the common element in these highlighted bins.

\begin{table*}[h!]
\centering \caption{Training overhead versus perfect BA probability, where $N_t = 32$, $N_r = 32$, and $M =
256$. As a comparison, the exhaustive search scheme requires $T=
262144$ time slots to complete the training process.}
\begin{tabular}{c|c|c}
\hline \hline
 \rowcolor[HTML]{D6DBDF} Parameters setting& Training overhead & Perfect Beam Alignment Probability\\
\hline \rowcolor[HTML]{F2F4F4} $R_{\rm BS} = 8,R_{\rm UE} = 4,
Q=16,L = 4$.
 &  $T = 2048 $ & $P =95.76\% $  \\
 \rowcolor[HTML]{E5E8E8} $R_{\rm BS} = 4,R_{\rm UE} = 4, Q=32,L = 4$. &  $T =2048 $ & $P = 97.00\% $  \\
\rowcolor[HTML]{F2F4F4} $R_{\rm BS} = 4,R_{\rm UE} = 4, Q=32,L =
5$.
 &  $T =2560 $ & $P = 99.64\% $  \\
\hline \hline
\end{tabular}
\label{table-comp}
\end{table*}

\subsection{Discussions}
Recall that the exhaustive search scheme has a sample complexity
of $N_t N_r M$. As a comparison, to achieve the same spatial
resolution, the sample complexity for the proposed
multi-directional BT scheme is $T=SL$. Based on the probability
theory, it is not difficult to derive the exact number of
batch-mode scanning rounds $L$ required for attaining a decent
probability that the best beam tuple can be correctly identified.
Specifically, it can be analyzed that the number of batch-mode
scanning rounds $L$ is in the order of
$\mathcal{O}(\log(\max\{N_t,N_r,M\}))$, which is usually much less
than $R_{\text{BS}} R_{\text{UE}} Q$. In Table \ref{table-comp},
an example is provided to show the amount of training overhead
required to attain perfect BA with a decent probability, from
which we see that it only takes the proposed method less than 1\%
of the training overhead required by the exhaustive search scheme
to achieve BA with a same spatial resolution.

We highlight some important advantages of our proposed method
compared to the hierarchical search scheme. First, unlike the
hierarchical search scheme, the proposed approach does not involve
multiple rounds of interactions between BS/RIS and the user, thus
can be straightforwardly extended to multi-user scenarios. Also,
compared with the hierarchical search scheme, the proposed method
is more robust to noise due to the use of narrow beams, which
helps increase the probability of identifying the best beam tuple
in the low SNR regime. For clarity, in Table \ref{table-Alg-comp},
we provide a concise overview of some key aspects of our proposed
and existing BT techniques.

\begin{table*}[h!]
\centering \caption{Advantages and limitations of different BT
methods}
\begin{threeparttable}
\begin{tabular}{ c|m{4cm}<{\centering}|m{4cm}<{\centering}|m{4cm}<{\centering}}
\hline \hline
{ \cellcolor[HTML]{D6DBDF} \diagbox{Features}{Algorithms}} &
\cellcolor[HTML]{D6EAF8} Exhaustive search & \cellcolor[HTML]{EBDEF0} Hierarchical search
&  \cellcolor[HTML]{FCF3CF} Multi-directional search\\
\hline \cellcolor[HTML]{EBEDEF} Training overhead  &
\cellcolor[HTML]{EBF5F8} $N_tMN_r$, extremely high overhead &
\cellcolor[HTML]{F5EEF8} $C^3 \log_C \max \{ N_t, M, N_r\}$, low
overhead
& \cellcolor[HTML]{FEF9E7} $N_t MN_rL/(R_{\rm BS}{R}_{\rm UE}Q)$, moderately low overhead   \\
 \hline
\cellcolor[HTML]{EBEDEF} Robustness against noise & \cellcolor[HTML]{EBF5F8}  Extremely high
& \cellcolor[HTML]{F5EEF8} Susceptible & \cellcolor[HTML]{FEF9E7} High  \\
\hline \cellcolor[HTML]{EBEDEF} Frequent feedback&
\cellcolor[HTML]{EBF5F8} Not required
&  \cellcolor[HTML]{F5EEF8}  Required & \cellcolor[HTML]{FEF9E7}  Not required \\
\hline \cellcolor[HTML]{EBEDEF} Extension to multi-user scenarios
 &  \cellcolor[HTML]{EBF5F8} Straightforward   &  \cellcolor[HTML]{F5EEF8}
 Require careful coordination
 & \cellcolor[HTML]{FEF9E7} Straightforward \\
\hline \hline
\end{tabular}
\begin{tablenotes}
\item $C$ denotes the number of beams examined in each layer.
\end{tablenotes}
\end{threeparttable}
\label{table-Alg-comp}
\end{table*}

\subsection{Performance Evaluation}
\begin{figure*}[!t]
\centering \subfigure[AAR versus SNR for LOS
scenarios.]{\includegraphics[width=3in]{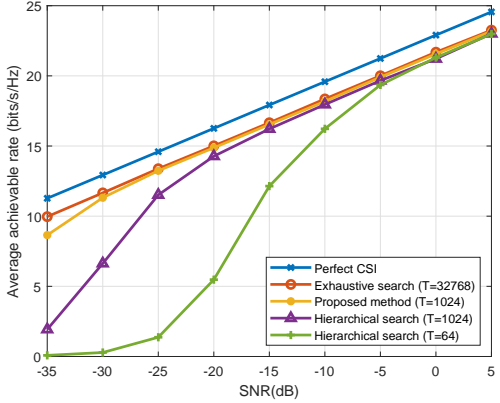}}\hfil
\subfigure[AAR versus SNR for NLOS
scenarios.]{\includegraphics[width=3in]{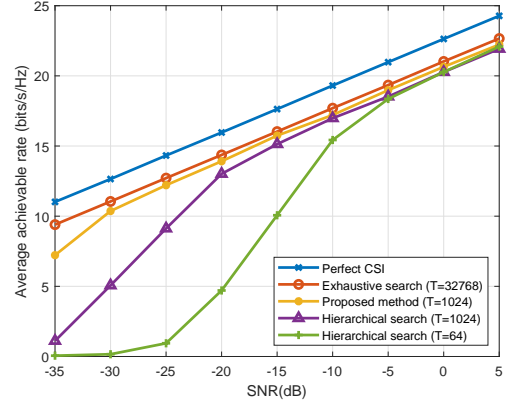}}
\caption{Average achievable rates (AARs) attained by respective BA
methods, where $N_t  = 128$, $N_r=1$, $M = 256$, $R_{\rm BS} = 8$,
$R_{\rm UE} = 1$ and $Q = 16$. The number of time slots for
training is set to $T=1024$ for our proposed method, and set to
$T=64$ and $T=1024$ for the hierarchical search scheme. The
training overhead for the exhaustive search scheme is $T= MN_tN_r
= 32768$.} \label{fig_simu}
\end{figure*}

Fig. \ref{fig_simu} depicts the spectral efficiency attained by
respective BT schemes. Results are averaged over $10^4$
independent runs. Also, we include the performance of the joint
beamforming method \cite{WangFang20} which assumes the perfect
knowledge of the full CSI. Clearly, this joint beamforming method
based on full CSI provides an upper bound on the achievable rate
attained by any BT schemes. The Rician factors for both the B-R
channel and the R-U channel are set to 13.2dB for the LOS
scenario, while for the NLOS scenario, we set the Rician factor of
the R-U channel to 0dB \cite{AkdenizLiu14}. It can be observed
that the spectral efficiency achieved by the exhaustive search
scheme is only slightly lower than that attained by assuming the
knowledge of full CSI. Another observation is that the proposed BT
method is able to obtain performance close to that of the
exhaustive search scheme with substantially fewer training
measurements. Specifically, the proposed method requires only
$3\%$ of the training overhead needed by the exhaustive search
scheme. For the hierarchical search scheme, it only needs
$T=2^3\times \log_2(M)=64$ time slots to complete the BT process,
which is the least among all three BT schemes. Nevertheless, due
to the use of wide beams at early stages, it performs poorly in
the low SNR regime. To make a fair comparison, we enlarge the
transmission time of the hierarchical search scheme such that its
training overhead is equal to that of our proposed method. Due to
the accumulated signal power, we see that the performance of the
hierarchical search scheme can be substantially improved.
Nevertheless, our proposed method still presents a clear
performance advantage over the hierarchical search scheme in the
low SNR regime. Such an advantage is highly desirable as it brings
an improve signal coverage for mmWave communications.

\section{Open Research Issues}
BT is essential for realizing the potential of RIS-assisted mmWave
systems. Despite some early studies on this subject, extensive
work is still needed to investigate this challenging problem from
both theoretical and practical aspects. In the sequel, we outline
several important open issues in BT for RIS-assisted mmWave
systems.

\subsection{Implementation Challenges}
There are several implementation challenges to be overcome for BT
for RIS-assisted mmWave systems. First, due to hardware
limitations, the phase shift coefficients at the RIS may only take
discrete values drawn from a finite set. As a result, the
generated beam based on coarsely quantized phase shifts deviates
from the ideal beam pattern and affects the accuracy of BA. It is
therefore of practical significance to analyze the impact of
hardware constraints/imperfections on the BT performance and
develop robust BT algorithms. Second, to achieve joint beam
training, the BS and the RIS need to be well synchronized. How to
achieve accurate synchronization between the BS and the RIS is
another important issue that should be addressed. Lastly, the
control link between the BS and the smart controller attached to
the RIS is capacity-limited. The control signal should be
suppressed as much as possible without sacrificing precise signal
transmission, which needs further investigation.

\subsection{Joint Localization and Beam Training}
RIS-assisted mmWave wireless systems offer great opportunities for
accurate localization and sensing due to its large bandwidth and
massive antenna array. Specifically, BT provides an estimate of
the AoD and AoA associated with the LOS/virtual LOS component
between the BS and the user. With the knowledge of the BS's and
RIS's locations, localization of the user is possible based on the
estimated angular parameters. On the other hand, the location
information and other situation awareness techniques can be
utilized to enhance the efficiency of the BT process. For example,
if the BS has some coarse information about the location of the
user, the angular space to be scanned at the BS/RIS can be
narrowed down. Accordingly, a more efficient BT sequence can be
devised to improve the BT efficiency by searching those areas of
interest instead of scanning the entire angular space.

\subsection{Extension To Multi-RIS Scenarios}
As a cost-effective and energy-efficient means for reconfiguring
the wireless propagation environment, it is desirable to deploy
multiple RISs to more effectively overcome blockage and improve
signal coverage for mmWave communications. Developing an efficient
BT method for multi-RIS-assisted mmWave systems is thus important
and meanwhile challenging. A straightforward approach is to
perform BT in an alternating manner, i.e. we sequentially turn on
one of RISs and switch off the rest RISs, and employ the
previously discussed BT methods to acquire the associated dominant
path's information. Such an approach, however, involves a sample
complexity (i.e. training overhead) that grows linearly with the
number of RISs. In this regard, it would be interesting to explore
a more efficient method that can jointly perform BT for multiple
BS-RIS-user links.

\section{Concluding Remarks}
In this article, we first discussed the extension of conventional
BT methods to RIS-assisted mmWave systems. The advantages and
limitations of each approach are elaborated. To address drawbacks
of existing methods, we presented a new multi-directional BT
method. Numerical results show that the proposed method can
achieve decent BA performance under low SNR environments, even
with a moderate amount of training overhead that is far less than
that of the exhaustive search scheme. Finally, several open
research issues for RIS-assisted BT are presented.

%\bibliography{newbib}

\begin{thebibliography}{10}
\providecommand{\url}[1]{#1} \csname url@rmstyle\endcsname
\providecommand{\newblock}{\relax}
\providecommand{\bibinfo}[2]{#2}
\providecommand\BIBentrySTDinterwordspacing{\spaceskip=0pt\relax}
\providecommand\BIBentryALTinterwordstretchfactor{4}
\providecommand\BIBentryALTinterwordspacing{\spaceskip=\fontdimen2\font
plus \BIBentryALTinterwordstretchfactor\fontdimen3\font minus
  \fontdimen4\font\relax}
\providecommand\BIBforeignlanguage[2]{{%
\expandafter\ifx\csname l@#1\endcsname\relax
\typeout{** WARNING: IEEEtran.bst: No hyphenation pattern has been}%
\typeout{** loaded for the language `#1'. Using the pattern for}%
\typeout{** the default language instead.}%
\else \language=\csname l@#1\endcsname \fi #2}}

\bibitem{AkdenizLiu14}
M.~R. Akdeniz, Y.~Liu, M.~K. Samimi, S.~Sun, S.~Rangan, T.~S.
Rappaport, and
  E.~Erkip, ``Millimeter wave channel modeling and cellular capacity
  evaluation,'' \emph{IEEE J. Sel. Areas Commun.}, vol.~32, no.~6, pp.
  1164--1179, Jun. 2014.

\bibitem{RappaportXing19}
T.~S. Rappaport, Y.~Xing, O.~Kanhere, S.~Ju, A.~Madanayake,
S.~Mandal,
  A.~Alkhateeb, and G.~C. Trichopoulos, ``Wireless communications and
  applications above 100 {GH}z: Opportunities and challenges for 6{G} and
  beyond,'' \emph{IEEE Access}, vol.~7, pp. 78\,729--78\,757, Jun. 2019.

\bibitem{AbariBharadia17}
O.~Abari, D.~Bharadia, A.~Duffield, and D.~Katabi, ``Enabling
high-quality
  untethered virtual reality,'' in \emph{Proc. 14th USENIX Symp. Netw. Syst.
  Des. Implement. (NSDI)}, Boston, MA, Mar. 27-29, 2017, pp. 531--544.

\bibitem{WangFang19}
P.~Wang, J.~Fang, X.~Yuan, Z.~Chen, and H.~Li, ``Intelligent
reflecting
  surface-assisted millimeter wave communications: Joint active and passive
  precoding design,'' \emph{IEEE Trans. Veh. Technol.}, vol.~69, no.~12, pp.
  14\,960--14\,973, Dec. 2020.

\bibitem{TanSun18}
X.~{Tan}, Z.~{Sun}, D.~{Koutsonikolas}, and J.~M. {Jornet},
``Enabling indoor
  mobile millimeter-wave networks based on smart reflect-arrays,'' in
  \emph{Proc. IEEE Int. Conf. Comput. Commun. (INFOCOM)}, Honolulu, Hawaii,
  Apr. 15-19 2018, pp. 270--278.

\bibitem{LiaskosNie18}
C.~{Liaskos}, S.~{Nie}, A.~{Tsioliaridou}, A.~{Pitsillides},
S.~{Ioannidis},
  and I.~{Akyildiz}, ``A new wireless communication paradigm through
  software-controlled metasurfaces,'' \emph{IEEE Commun. Mag.}, vol.~56, no.~9,
  pp. 162--169, Sept. 2018.

\bibitem{WuZhang21}
Q.~Wu, S.~Zhang, B.~Zheng, C.~You, and R.~Zhang, ``Intelligent
reflecting
  surface-aided wireless communications: A tutorial,'' \emph{IEEE Trans.
  Commun.}, vol.~69, no.~5, pp. 3313--3351, 2021.

\bibitem{WangFang20}
P.~Wang, J.~Fang, H.~Duan, and H.~Li, ``Compressed channel
estimation for
  intelligent reflecting surface-assisted millimeter wave systems,'' \emph{IEEE
  Signal Process. Lett.}, vol.~27, pp. 905--909, May 2020.

\bibitem{WeiShen21}
X.~Wei, D.~Shen, and L.~Dai, ``Channel estimation for {RIS}
assisted wireless
  communications -- part {II}: An improved solution based on double-structured
  sparsity,'' \emph{IEEE Commun. Lett.}, vol.~25, no.~5, pp. 1403--1407, Jan.
  2021.

\bibitem{WanGao21}
Z.~Wan, Z.~Gao, F.~Gao, M.~D. Renzo, and M.-S. Alouini,
``Terahertz massive
  {MIMO} with holographic reconfigurable intelligent surfaces,'' \emph{IEEE
  Trans. Commun.}, vol.~69, no.~7, pp. 4732--4750, Mar. 2021.

\bibitem{YouZheng20}
C.~You, B.~Zheng, and R.~Zhang, ``Fast beam training for
{IRS}-assisted
  multiuser communications,'' \emph{IEEE Wireless. Commun. Lett.}, vol.~9,
  no.~11, pp. 1845--1849, Nov. 2020.

\bibitem{XiaoHe16}
Z.~Xiao, T.~He, P.~Xia, and X.-G. Xia, ``Hierarchical codebook
design for
  beamforming training in millimeter-wave communication,'' \emph{IEEE Trans.
  Wireless Commun.}, vol.~15, no.~5, pp. 3380--3392, May. 2016.

\bibitem{HengAndrews21}
Y.~Heng and J.~G. Andrews, ``Machine learning-assisted beam
alignment for
  mmwave systems,'' in \emph{Proc. IEEE Global Commun. Conf. (GLOBECOM)},
  Waikoloa, HI, USA, 9-13, Dec. 2019, pp. 1--6.

\bibitem{WeiZhou17}
T.~Wei, A.~Zhou, and X.~Zhang, ``Facilitating robust 60 {GH}z
network
  deployment by sensing ambient reflectors,'' in \emph{Proc. 14th USENIX Symp.
  Netw. Syst. Des. Implement. (NSDI)}, Boston, MA, Mar. 27-29, 2017, pp.
  213--226.

\bibitem{WangFang21a}
P.~Wang, J.~Fang, W.~Zhang, and H.~Li, ``Fast beam training and
alignment for
  {IRS}-assisted millimeter wave/terahertz systems,'' \emph{IEEE Trans.
  Wireless Commun.}, Early Access, 2021, DOI: 10.1109/TWC.2021.3115152.

\end{thebibliography}
%\bibliographystyle{IEEEtran}

\end{document}